\documentclass[preprint2]{aastex6}

\usepackage{graphicx,epsfig,fancyhdr,psfig,rotating,amsmath,epsf,txfonts,natbib,epstopdf,multirow}
\def \kms {{\rm km\;s$^{-1}$}}

\def \arcsec {$^{''}$}
\def \siiv {Si\,{\sc iv}}

\def \datai {{DATA1}}
\def \dataii {{DATA2}}
\newcommand{\mathd}{\ensuremath{{\rm d}}}
\graphicspath{{./newfigs/}}

\begin{document}
\title{Observations of upward propagating waves in the transition region and corona above Sunspots}
\author{
\sc{Zhenyong Hou, Zhenghua Huang, Lidong Xia, Bo Li, Hui Fu}
}
\affil{Shandong Provincial Key Laboratory of Optical Astronomy and Solar-Terrestrial Environment, Institute of Space Sciences,
Shandong University, Weihai, 264209 Shandong, China; {\it xld@sdu.edu.cn}}
\begin{abstract}
We present observations of persistent oscillations of some bright features in the upper-chromosphere/transition-region above sunspots taken by IRIS SJ 1400\,\AA\ and  upward propagating quasi-periodic disturbances along coronal loops rooted in the same region taken by AIA 171\,\AA\ passband.
The oscillations of the features are cyclic oscillatory motions without any obvious damping.
The amplitudes of the spatial displacements of the oscillations are about 1\arcsec.
The apparent velocities of the oscillations are comparable to the sound speed in the chromosphere,
but the upward motions are slightly larger than that of the downward.
The intensity variations can take 24--53\% of the background, suggesting nonlinearity of the oscillations.
The FFT power spectra of the oscillations show dominant peak at a period of about 3 minutes, in consistent with the omnipresent 3 minute oscillations in sunspots.
The amplitudes of the intensity variations of the upward propagating coronal disturbances
are 10--15\% of the background.
The coronal disturbances have a period of about 3 minutes, and
propagate upward along the coronal loops with apparent velocities in a range of 30$\sim$80\,\kms.
We propose a scenario that the observed transition region oscillations are powered continuously by upward propagating shocks, and
the upward propagating coronal disturbances can be the recurrent plasma flows driven by shocks or responses of degenerated shocks that become slow magnetic-acoustic waves after heating the plasma in the coronal loops at their transition-region bases.

\end{abstract}
\keywords{Sunspots - Sun: transition region - Sun: corona - Sun:oscillations - Methods: observational}

\maketitle

\section{Introduction}
\label{sect_intro}
As one of the prominent phenomena observed in sunspots, oscillation has been intensively studied in the past\,\citep{2003A&ARv..11..153S,2015LRSP...12....6K}.
The sunspot oscillations offer a powerful tool to probe the structure of magnetic flux tubes beneath sunspot photosphere\,\citep{1985AuJPh..38..811T} and possibly heat the chromosphere and corona above\,\citep{2015LRSP...12....6K}.
The periods of sunspot oscillations are dominant in the bands of 5 minutes and 3 minutes.
Their footprints have been found throughout the solar atmosphere above sunspots, from the photosphere to the corona.
More details on sunspot oscillations can be found in published reviews\,\citep[e.g.][]{1985AuJPh..38..811T,2006RSPTA.364..313B,2015LRSP...12....6K}

\par
The 5 minute sunspot oscillations are mostly found at the photospheric level. They are the response to the widespread 5 minute $p$-mode oscillations in the surrounding quiet photosphere\,\citep{1985AuJPh..38..811T}.
Their power in sunspots is also much smaller than that in the surrounding quiet photosphere\,\citep{1968SoPh....4..286H,1976A&A....50..367S, 2007PASJ...59S.631N}.
They are normally identified as amplitude fluctuations in line-of-sight (LOS) velocity measured by the photospheric spectral lines, such as Fe\,{\sc i}, Ti\,{\sc i} and TiO\,\citep{1968SoPh....4..286H,1972SoPh...27...80B,1976A&A....50..367S,1984SoPh...94...99B,1985ApJ...294..682L,1990Ap&SS.170..121A,1997AN....318..129L}.
In the photospheric level, the 5 minute oscillations are dominated in the outer penumbra with a ring of minimum power halfway between the inner and the outer penumbral boundary\,\citep{1988ApJ...334.1054L,1990SoPh..125...31B}. Via both velocity and intensity variations, 5 minute oscillations can also be found in the chromosphere of sunspots, especially dominant in the penumbrae\,\citep{1985ApJ...294..682L,1997A&A...324..743S,2007PASJ...59S.631N,2014ApJ...792...41Y}.
The 5 minute oscillations in the umbrae are coherent between the the photosphere and low chromosphere, with a small, positive phase difference\,\citep{1985ApJ...294..682L}.

\par
As another common oscillation mode in sunspots, 3 minute oscillation is firstly discovered in the intensity disturbance of the emission of Ca\,{\sc ii}\,K from the umbrae and named ``umbral flashes''\,\citep{1969SoPh....7..351B}.
Most observations of 3 minute oscillations have been reported with the variations of brightness of the chromospheric Ca\,{\sc ii}\,H and K lines and velocities derived from Ca\,{\sc ii}\,H and K lines and He\,{\sc ii}\,10830\,\AA\ triplet.
Although the 3 minute oscillations can also be found in any atmospheric levels,
they are the dominant mode of oscillations in the chromosphere, transition region and corona above sunspot umbrae\,\citep[e.g.][]{1982ApJ...253..939G, 1987SoPh..108...61G, 1987ApJ...312..457T, 1995A&A...300..539K, 2002A&A...387L..13D, 2002SoPh..207..259B, 2004SoPh..221..237B, 2003A&A...403..277R, 2006ApJ...640.1153C,2016ApJ...816...30S,2016ApJ...817..117S}.
The Hinode/SOT data of chromosphere have also revealed that the 3 minutes oscillations are dominant in the umbrae while the 5 minute oscillations are concentrating in the penumbrae\,\citep[e.g.][]{2007PASJ...59S.631N,2014ApJ...792...41Y,2014A&A...561A..19Y}.
Since the 3 minute oscillations are often well distinct from the 5 minute oscillations in the power spectra without any correlations, they are believed to be a different oscillation mode in sunspots\,\citep[e.g.][]{1976A&A....49..463S,1985ApJ...294..682L,1986ApJ...301..992L,1987SoPh..108...61G,2006ApJ...640.1153C}.
Recently, there is a report on a single case that the 3 minute oscillation in the sunspot chromosphere was excited by a strong downflow event in a sunspot\,\citep{2016ApJ...821L..30K}.

\par
Many observations of the 3 minute oscillations in the transition region and corona have been reported.
Their footprints in the transition regions above sunspots have been found in the intensity and velocity temporal variations of the transition region spectral lines\,\citep[see e.g.][]{1982ApJ...253..939G,1987ApJ...312..457T}. Recently, IRIS observations have revealed clusters of jet-like events in the transition region above light bridges of sunspots with a repeating period of 3--4 minutes, suggesting that the oscillation-related events might provide energy to heat the upper atmosphere of the sunspots\,\citep{2015ApJ...804L..27Y, 2015MNRAS.452L..16B, 2016A&A...589L...7H,2016ApJ...833L..18Y,2017ApJ...843L..15Y}.
With the imaging data of the corona, it has been frequently reported that the 3 minute oscillations are found in the coronal loops rooted in the sunspot umbral regions\,\citep[e.g.][]{2000A&A...362.1151N, 2002A&A...387L..13D,2002SoPh..207..259B, 2004SoPh..221..237B} and flare loops\,\citep{2017arXiv170910059L}.

\par
While the 3 minute oscillations are found in the transition region and corona, their physical nature and their connections to the chromospheric oscillations have been intensively studied in the past.
By studying 3 minute oscillations in sunspots, \citet{1978SoPh...58..347G} reported a 12\,s delay in the intensity of H$\alpha$ line center to that in the wing of H$\alpha$ at $\pm$\,0.39\,\AA, and they interpret this as the results of upward propagating waves.
\citet{1982ApJ...253..939G} found that the maximum intensity was in phase with maximum blueshift in C\,{\sc iv} emission of the 3 minute sunspot oscillations, and this supports that the 3 minute sunspot oscillations in the transition region are responses of upward propagating acoustic waves.
The upward propagating wave scenario has also supported by later studies\,\citep[e.g.][]{1983A&A...123..263U,1984ApJ...277..874L,1984SoPh...91...33H,1999ApJ...517L.159B,1999ApJ...511L.121B,2000SoPh..191..129B,
  2001ApJ...552L..77B,2002SoPh..207..259B,2004SoPh..221..237B,2001A&A...373L...1M,2006ApJ...640.1153C,
  2012ApJ...746..119R}.
Upward acoustic waves are subjected to cut-off frequencies,
which are determined by the inclination of the local magnetic field\,\citep[see][and references therein]{2014A&A...561A..19Y,2016ApJS..224...30Y}.
By investigating oscillations with frequencies in the range of 5--9\,mHz (about 2--3 minutes in period) above sunspots, \citet{2012ApJ...746..119R} found that high-frequency oscillations are more pronounced in the central region of the umbrae while the lower frequency ones are prominent at the outer regions, and
they suggest that the reason could be a variation of the magnetic field inclination across the umbra at the level of temperature minimum.

\par
The upward propagating waves in sunspots may show nonlinear properties.
\citet{1984ApJ...277..874L} reported that the 3 minute oscillations in sunspot umbra were upward-propagating acoustic (or slow mode) disturbances that become nonlinear and develop into shock waves in the upper layers.
Many observations have confirmed the nonlinearity and signatures of shock waves of the 3 minute upward propagating waves in the chromosphere\,\citep[e.g.][]{1999ApJ...517L.159B,2003A&A...403..277R,2006ApJ...640.1153C,2015ApJ...805L..21C}.
With imaging and spectral data of sunspot oscillations taken by IRIS, \citet{2014ApJ...786..137T} reported on the first direct evidence of shock waves in the transition region above sunspots. \citet{2014ApJ...786..137T} also found that the transition region oscillation (seen in Si\,{\sc iv}) lags the chromospheric ones (seen in C\,{\sc ii} and Mg\,{\sc ii}), indicating their propagating property. Signatures of shock waves in the transition region have also been found above light bridge\,\citep{2017ApJ...838....2Z}.

\par
In the present work, we report on two sets of observations that show responses to upward propagating waves in the sunspots while they are propagating as intensity disturbances along coronal loops and while they are passing structures with enhanced emission in the transition region. These structures present as bright features seemingly floating above the dark sunspot umbrae.
Unlike the signatures of oscillations revealed in the light-curves of a particular location, the transition region structures here present persistent oscillations with clear spatial displacements in the high-resolution IRIS data.
In what follows, we describe our observations in Section\,\ref{sect_obs},
present the results in Sections\,\ref{sect_res1}--\ref{sect_link},
make the discussion in Section\,\ref{sect_disc}
and summarize our findings in Section\,\ref{sect_concl}.

\begin{figure*}[!ht]
\centering
\includegraphics[trim=0cm 0cm 0.5cm 0cm,clip,width=0.95\textwidth]{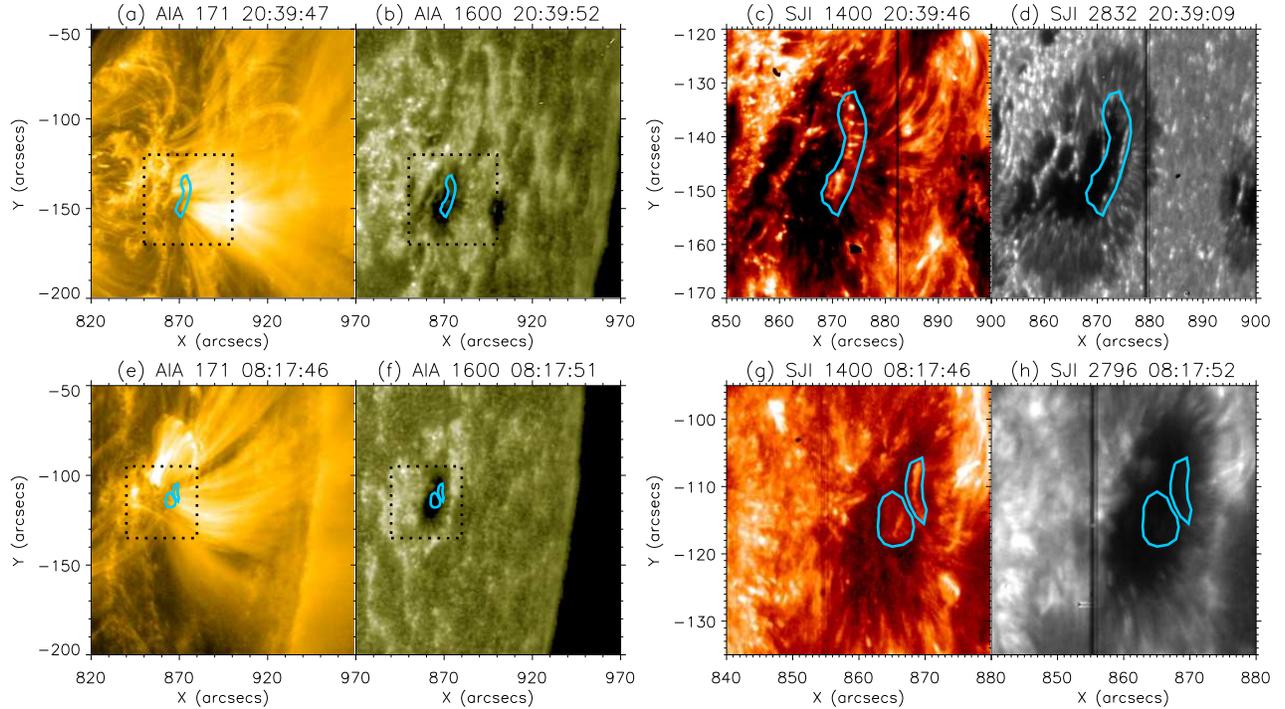}
\caption{
The regions of interest studied in the present work.
Panels (a)--(d) show the region in \datai:
(a) the zoomed-out field of view giving the coronal environment with large scale loops seen in AIA\,171\,\AA;
(b) the zoomed-out field of view displaying the chromospheric environment with sunspots and plage regions seen in AIA\,1600\,\AA;
(c) the zoom-in view of the region of interest in IRIS SJ 1400\,\AA\ observations, and the region is
enclosed by black dotted lines in panels (a)\&(b).
(d) the same field of view of the region as panel (c), but in IRIS SJ 2832\,\AA\ observations giving the detailed view of the sunspots in chromospheric emission.
The oscillating feature seen in IRIS SJ 1400\,\AA\ is denoted by the contour painted in cyan.
Panels (e)--(h) present the same as panels (a)--(d), but for the region of interest in \dataii, also panel (h) presents the region seen in IRIS SJ 2796\,\AA\ passband.
}
\label{f1}
\end{figure*}

\begin{figure*}[!ht]
\centering
\includegraphics[trim=0cm 0cm 0cm 0.cm,width=0.9\textwidth]{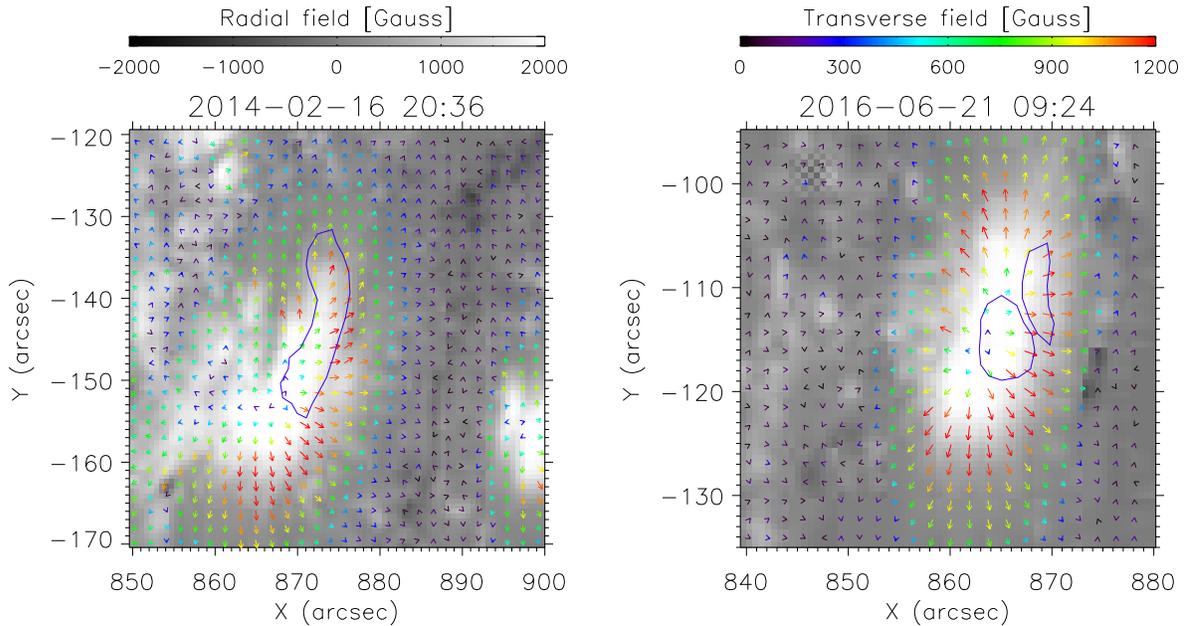}
\caption{
The phtospheric vector magnetic fields of the regions taken with HMI. The background images present the radial component (B$_r$) of the field while the transverse component is presented as colour-coded arrows.
The oscillating features are outlined by blue contour lines.
}
\label{fig_vmag}
\end{figure*}

\section{Observations}
\label{sect_obs}
The observations analysed in this study were taken by the Interface Region Imaging Spectrograph\,\citep[IRIS,][]{2014SoPh..289.2733D} and the Atmospheric Imager Assembly\,\citep[AIA,][]{2012SoPh..275...17L}.
Two sets of data are anlysed, including one taken on 2014 February 16 from 20:19\,UT to 21:04\,UT (hereafter, DATA1) and the other one taken on 2016 June 21 from 07:38\,UT to 10:01\,UT (hereafter, DATA2).

\par
The IRIS spectral data were taken in the very dense raster mode for \datai\ and medium sparse 2-step raster mode for \dataii.
They are not suitable for studying the high frequency phenomena, and only the time series of slit-jaw (SJ) images are used.
For \datai, the SJ images were taken in 1330\,\AA, 1400\,\AA\ and 2832\,\AA\ passbands with cadences of about 14\,s, 14\,s and 32\,s, respectively.
For \dataii, the SJ passbands of 1400\,\AA\ and 2796\,\AA\ were used and the data were taken with a cadence of about 11\,s.

\par
For AIA observations, the data taken by the EUV channel of 171\,\AA\ with a cadence of 12\,s are mainly used in this study. The temperature response of this channel peaks at 630\,000\,K, representative of emissions from the corona.

\par
The fields of view of \datai\ and \dataii\ are shown in Fig.\,\ref{f1}.
Both datasets were targeting at active regions approaching the limb, which allows better visibility of the radial displacement in the transition region and corona.
IRIS was targeting at NOAA 11974 while taking \datai\, and at NOAA 12553 while taking \dataii.

\par
We use the HMI\,\citep{Schou2012hmi} Full-Disk disambiguated vector magnetic field observations\,\citep{2014SoPh..289.3483H}, i.e. `hmi.B\_720s' data series, to investigated the magnetic configurations of the sunspots. The HMI vector field has been obtained through the Very Fast Inversion of Stokes Vector\,\citep[VFISV,][]{ 2011SoPh..273..267B}, which is a Milne-Eddington based algorithm. A minimum energy method\,\citep{1994SoPh..155..235M, 2009SoPh..260...83L} is used to resolve the 180$^{\circ}$ ambiguity in the transverse field. The HMI instrument team provides four fits files, including field strength, inclination, azimuth and disambiguated data. The information in disambiguated data indicates whether the azimuth should be increased by 180 degrees. This can be processed by the \textit{ssw} procedure \textit{hmi\_disambig.pro} provided by the instrument team. We then obtain the radial component (B$_r$) and the transverse components of the photospheric vector magnetic field via \textit{ssw} procedure \textit{hmi\_b2ptr.pro}, and show in Fig.\,\ref{fig_vmag}.

\par
We aligned the IRIS SJ 1400\,\AA\ to the AIA 1600\,\AA\ images due to their strong continuum contribution, and the AIA 171\,\AA\ images can be then aligned with the 1600\,\AA\ through using images from the other AIA channels. In order to compare the time series obtained by different instruments, both IRIS SJ and AIA\,171\,\AA\ images were interpolated to a uniform time grid with 14\,s cadence for  \datai, and 12\,s cadence for \dataii.

\begin{figure*}[!ht]
\centering
\includegraphics[trim=1cm .5cm 1cm 0.cm,width=0.98\textwidth]{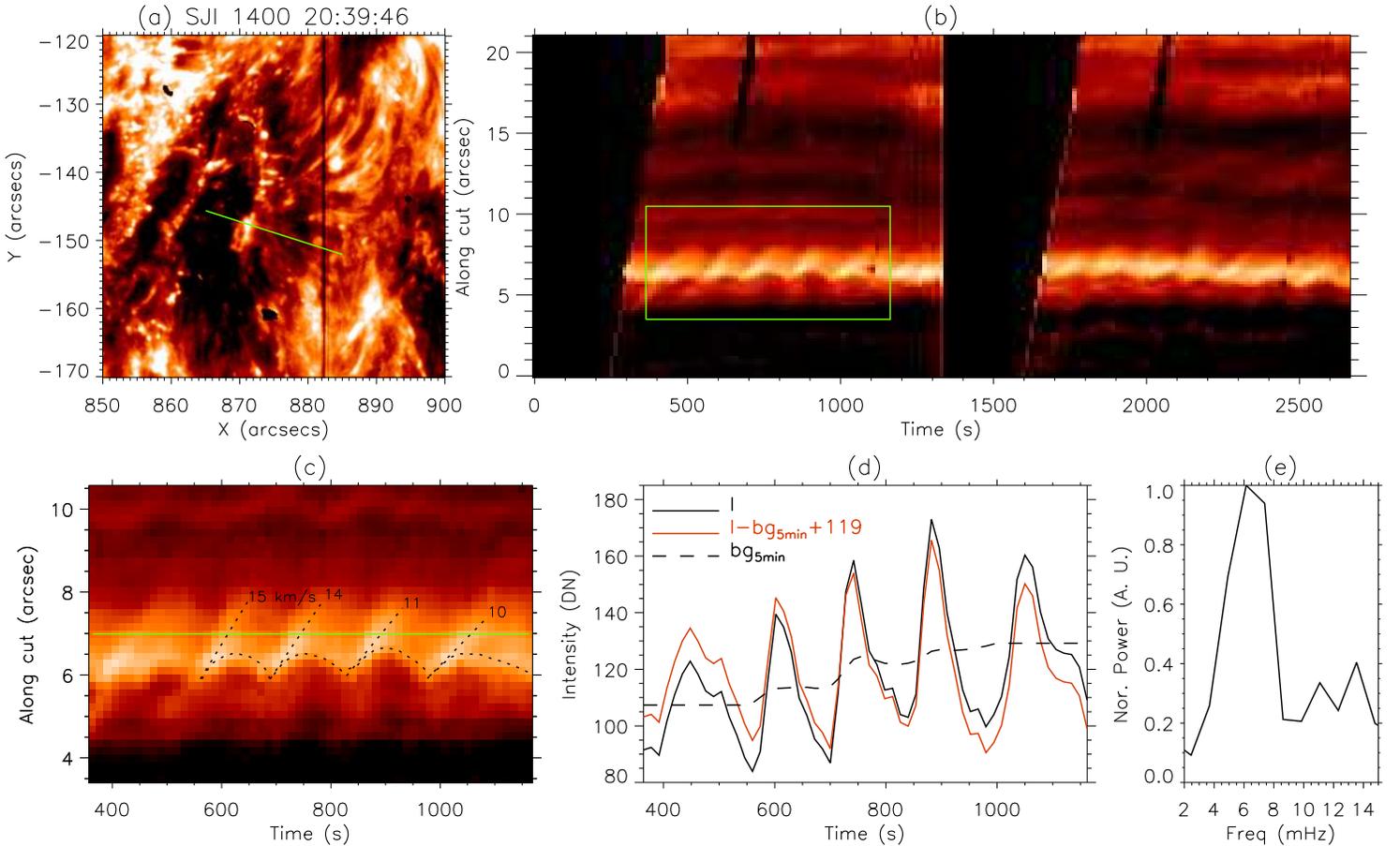}
\caption{
Analysis of the oscillations of the feature in \datai\ seen by IRIS SJ 1400\,\AA.
(a): The region viewed in IRIS SJ 1400\,\AA. The green solid line denotes the cut that we obtained the time-distance image.
(b): The time-distance image along the cut denoted in (a). The green rectangle denotes the certain time period and spatial range that are analysed in detail (panels c--e).
(c): The zoomed-in view of the region on the time-distance image shown in panel (b). The dotted lines denote the tracks of the oscillating feature, and the apparent speeds determined by the slopes of the upward-propagating tracks are marked (in units of \kms). The green line marks the location of the cut, whose light curve is shown in (d).
(d): The light curve. The original intensity variation is shown in black solid line; the background emissions obtained from 5-minute running means are shown in black dashed line; the relative intensity variation, i.e. the original intensities subtracted by the running means, is given in red solid line. A value of 119 (DN) has added to the relative intensity variation, in order to have a direct comparison to the original signal.
(e) The FFT power spectrum of the detrended light curve shown in panel (d).
}
\label{fig:os1:1400}
\end{figure*}

\section{Results from \datai }
\label{sect_res1}
In \datai, we pay attention on an arch bright feature with a size $\sim$18\arcsec\ extending south-north above the leading sunspot of AR11974 observed by the IRIS SJ 1400\,\AA\ channel (see Fig.\ref{f1}a--d).
Some jet-like features are coincidently appear in the same region, but they seem to be phenomena in the background and/or foreground.
By aligning the IRIS SJ 1400\,\AA\ image with SJ 2832\,\AA\ image, we found that
this arch feature is not associated with any light bridge in the sunspot.
It appears to be  bright feature floating above the dark sunspot umbra.
The photospheric vector magnetic field of the region is given in left panel of Fig.\,\ref{fig_vmag}.
In the region surrounding the feature, beside the single polarity radial component, we can also see strong transverse component, which might provides necessary magnetic tensions to support the feature floating above the sunspot umbra.
The oscillating feature was also appeared to be bright in the transition region \siiv\ emission as studied previously\,\citep[see the event 2 in][]{2016ApJ...829L..30H}.
Therefore, the feature should be considered as upper chromosphere and transition region phenomena.
The feature was persistently swaying at the east-west direction (see the animation given along with Fig.\,\ref{fig:os1:1400}).
While cross-checking with the AIA 171\,\AA\ images, we found that this feature was located in the footpoint region of a set of large coronal loops (see Fig.\,\ref{f1}). In AIA 171\,\AA\ images, we can observe that recurring bright disturbances are propagating along the set of large loops in a manner of quasi-periodicity (see the animation associated with Fig.\ref{fig:os1:1400}).

\subsection{Oscillating feature in the chromosphere/transition region}
\label{sect_trosc1}
In Fig.\,\ref{fig:os1:1400} and the associated animation, we present the evolution of the oscillating feature observed by IRIS SJ 1400\,\AA.
The feature appears to be oscillating near an equilibrium position, without any clear propagation along its length, indicating that the feature is perturbed in a coherent way.
To investigate the oscillating behaviour of the feature, we take a cut along the oscillating direction and along the coronal loops rooting in the region seen in AIA 171\,\AA\ passband. The time-distance image along the cut is given in Fig.\,\ref{fig:os1:1400}b, and two remarkable patterns can be seen on the time-distance image.

\par
The first remarkable pattern on the time-distance image is the recurring arch-shaped tracks (see the patterns marked by black dotted lines in Fig.\,\ref{fig:os1:1400}c).
These tracks are consistent with the motion of the bright transition region feature that is moving back and forth in a repetitive manner.
The apparent speeds of the raising (westward/upward) motions seem to be larger than that of falling-back.
The spatial displacement (i.e. amplitude) of the oscillation of the feature is $\lesssim$1\arcsec that can be well resolved in the high resolution IRIS SJ images and seen from the time-distance image (see Fig.\,\ref{fig:os1:1400}c).

\par
The other remarkable pattern is the branches of tracks that show unidirectional propagation (see the stripe-like patterns marked by dotted lines with speeds denoted in Fig.\,\ref{fig:os1:1400}c),
which were upward propagating along the cut (i.e. the coronal loop).
The slopes of these branches are almost identical to that of the raising stage of the oscillating feature.
The time-distance image also shows that the upward propagating component was separated when the main oscillating feature starts to decelerate and fall back.
This can also be seen by following the evolution of the oscillating feature using the animation attached to Fig.\,\ref{fig:os1:1400}.
Using the slopes of the stripe-like patterns, the apparent velocities of the upward propagating components are found to be in the range of 10$\sim$15\,\kms.
These values are comparable to the sound speed ($\sim10$\,\kms) in the average chromosphere,
but smaller than the Alfv\'en speed\,\citep[$\sim100$\,\kms, see][]{2007Sci...318.1574D}.

\par
We further analyzed a section of the time series that best shows periodicity of the phenomena (see Fig.\,\ref{fig:os1:1400}b).
The zoomed-in time-distance image along the cut taken during this period of time is given in Fig.\,\ref{fig:os1:1400}c. The light curve at the location marked by green line in Fig.\,\ref{fig:os1:1400}c is given in Fig.\,\ref{fig:os1:1400}d and used to analyze the periodicity.
The trend of the background emission was obtained as running means with a width of 5 minutes.
We found that the intensity amplitudes of the light-curve fluctuation take 24--37\% of the background, hinting at nonlinearity\,\citep{2014masu.book.....P,2014ApJ...786..137T}.
The amplitudes of the oscillations do not show any clear damping during the observations.

\par
In order to obtain the periodicity of oscillating phenomenon,
we first subtracted the trend from the light curve and
then applied the Fast Fourier Transform (FFT) to the de-trended light curve.
The power spectrum of the de-trended light curve is shown in Fig.\,\ref{fig:os1:1400}e.
It clearly shows that the power spectrum mainly peaks at about 6.2\,mHz, i.e. 2.7\,minutes in period.
This frequency (period) is consistent with the prominent 3 minute oscillation in sunspots.

\begin{figure*}[!ht]
\centering
\includegraphics[trim=1cm 0cm 1cm 0.cm,width=0.95\textwidth]{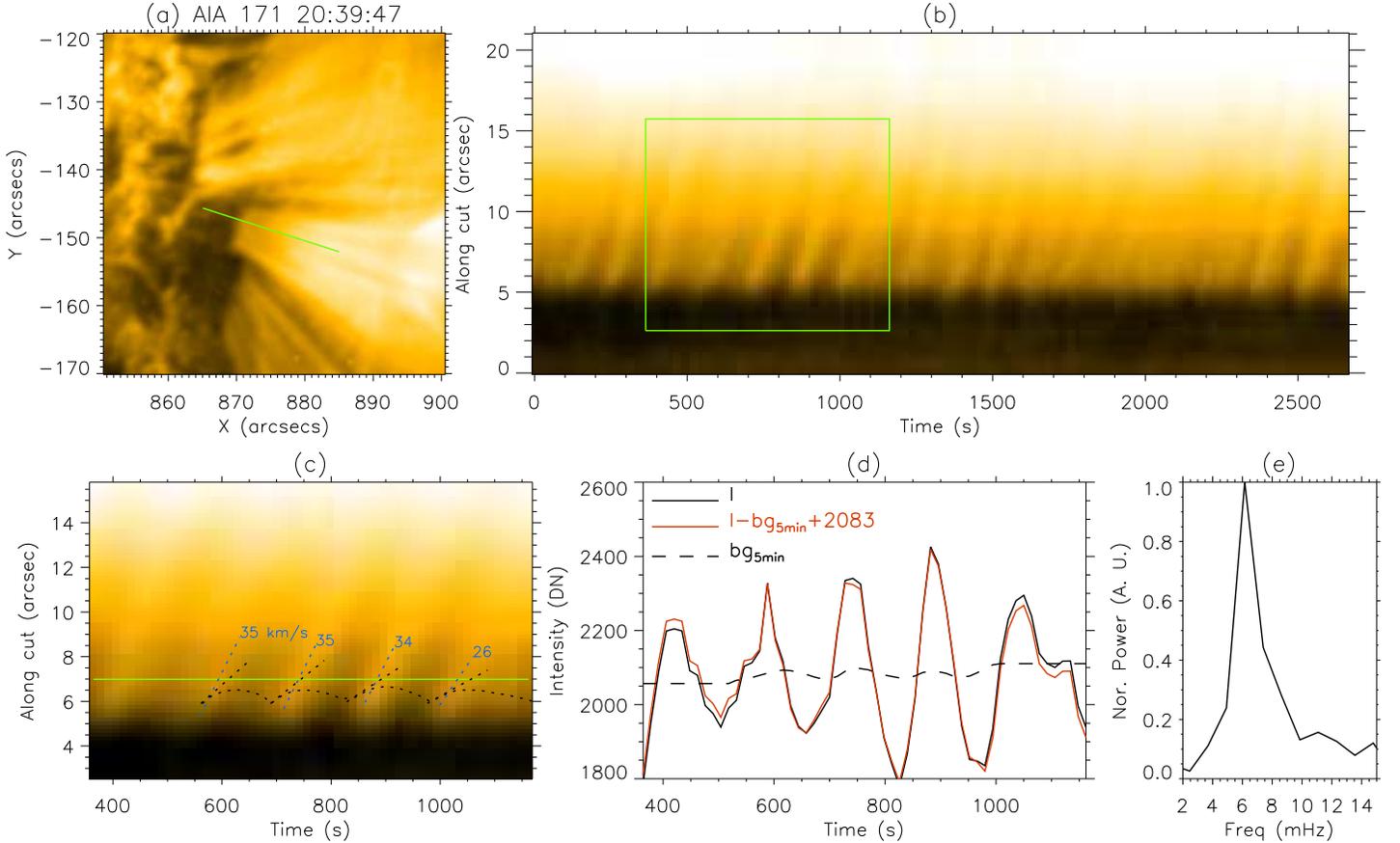}
\caption{
As Fig.\,\ref{fig:os1:1400}, but for the observations taken by AIA 171\,\AA, in which the upward propagating disturbances can be seen.
The cut marked as green line in panel (a) is identical to that used in the analysis of IRIS SJ 1400\,\AA\ as was shown in Fig.\,\ref{fig:os1:1400}.
The tracks of the upward propagating coronal disturbances are denoted by the cyan dotted lines with the marks of their apparent velocities (in units of \kms).
The black dotted lines in panel (c) mark the tracks of the oscillations of the feature in IRIS SJ 1400\,\AA\ (Fig.\,\ref{fig:os1:1400}).
}
\label{fig:os1:171}
\end{figure*}

\subsection{Upward propagating disturbance along coronal loops}
\label{sect_coosc1}
The periodic upward propagating coronal disturbances can be clearly seen in the AIA 171\,\AA\ images (Fig.\,\ref{fig:os1:171} and the online animation).
The time-distance image shown in Fig.\,\ref{fig:os1:171} was taken along the cut that is identical to the one used in IRIS SJ 1400\,\AA\ images.
The most pronounced feature seen on the time-distance image is the stripe-like patterns.
In Fig.\,\ref{fig:os1:171}c, we present the zoomed-in view of time-distance image taken from the period of time identical to that shown in Fig.\,\ref{fig:os1:1400}c.
The apparent velocities of the upward propagating disturbances are about 30\,\kms\ as determined from the slopes of the stripe-like tracks on the time-distance image, much larger than that determined in IRIS SJ images (see Fig.\,\ref{fig:os1:171}c).
If the projection effect is taken into account, the values could be larger.
These speeds are much smaller than the sound speed in the average corona ($\sim$120\,\kms).
Therefore, they are considerably subsonic even taking into account the projection effect.

\par
At the same location that used to produce the IRIS SJ light curve in Fig.\,\ref{fig:os1:1400}d,
the AIA 171\,\AA\ light curve was also generated and shown in Fig.\,\ref{fig:os1:171}d.
Again, the trend of the light curve was obtained as running means with a 5 minute period.
The amplitudes of the intensity fluctuations in the AIA 171\,\AA\ light curve are found to take about 15\% of the background.
This value is much smaller than that in the IRIS SJ 1400\,\AA\ observations.
The small amplitudes of the intensity fluctuation implies
that the nonlinear effect might not present in these propagating disturbances.
The FFT power spectrum of the AIA 171\,\AA\ light curve (Fig.\,\ref{fig:os1:171}) mainly peaks at 6.2\,mHz (i.e. 2.7 minutes in period),
which is consistent with that revealed by the IRIS SJ 1400\,\AA\ time series.
This indicates that both the propagating disturbances in the corona and the oscillation of the chromosphere/transition-region feature are closely associated with the 3 minute oscillations above the sunspot.

\begin{figure*}[!ht]
\centering
\includegraphics[trim=1cm 0cm 1cm 0.cm,width=0.98\textwidth]{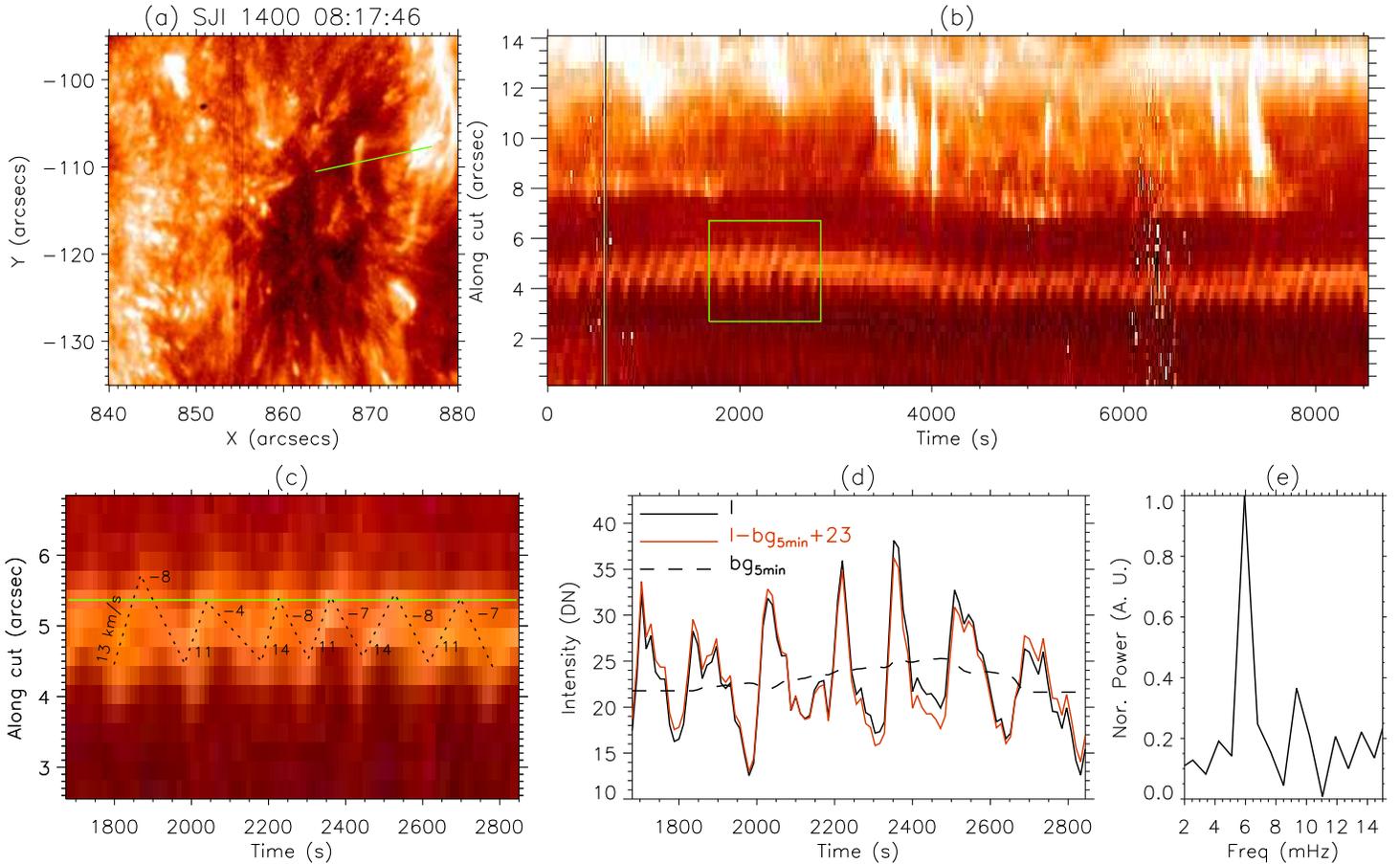}
\caption{
As Fig.\,\ref{fig:os1:1400}, but for the case observed in \dataii. In panel (c), the apparent velocities of the upward motions of the oscillations are given in positive values while the downward velocities are in negative values.
}
\label{fig:os2:1400}
\end{figure*}

\section{Results from \dataii}
\label{sect_res2}
Such phenomena present in \datai\ are not rare in the solar atmosphere.
Similarly, with IRIS SJ 1400\,\AA\ images of \dataii, we also observed some oscillating features above the umbra of the sunspot in AR12553 (see the animation along with Fig.\,\ref{fig:os2:1400}).
The features consist of two elongated bright cores surrounded by fuzzy patterns.
The features extended at the south-north direction, while the swaying motions of the features were perpendicular to their extending direction (see the animation with Fig.\,\ref{fig:os2:1400}).
As same as that shown in \datai, the features do not show any visible response in the chromosphere as seen in IRIS SJ 2796\,\AA\ (see Fig.\,\ref{f1}).
They seem to be bright structures in the transition region (or upper chromosphere) floatingp above the dark sunspot umbra (see Fig.\,\ref{f1} and the animation associated with Fig.\,\ref{fig:os2:1400}).
The photospheric vector magnetic field of the region is given in right panel of Fig.\,\ref{fig_vmag},
which shows very similar magnetic environment as the feature in \datai, i.e., having very strong transverse magnetic field.
Compare to the feature presented in \datai, these features appear to be cleanerp without any jet-like features in the surrounding.
The AIA 171\,\AA\ images reveal that many large coronal loops are rooted in the same region  and extending from east to west (see Fig.\,\ref{fig:os2:171}). Quasi-periodic disturbances are found to propagate along these coronal loops (see the animation associated with Fig.\,\ref{fig:os2:1400}).

\subsection{Oscillating features seen in IRIS SJ observations}
\label{sect_trosc2}
\begin{figure*}[!ht]
\centering
\includegraphics[trim=1cm .5cm 1cm 0.cm,width=0.9\textwidth]{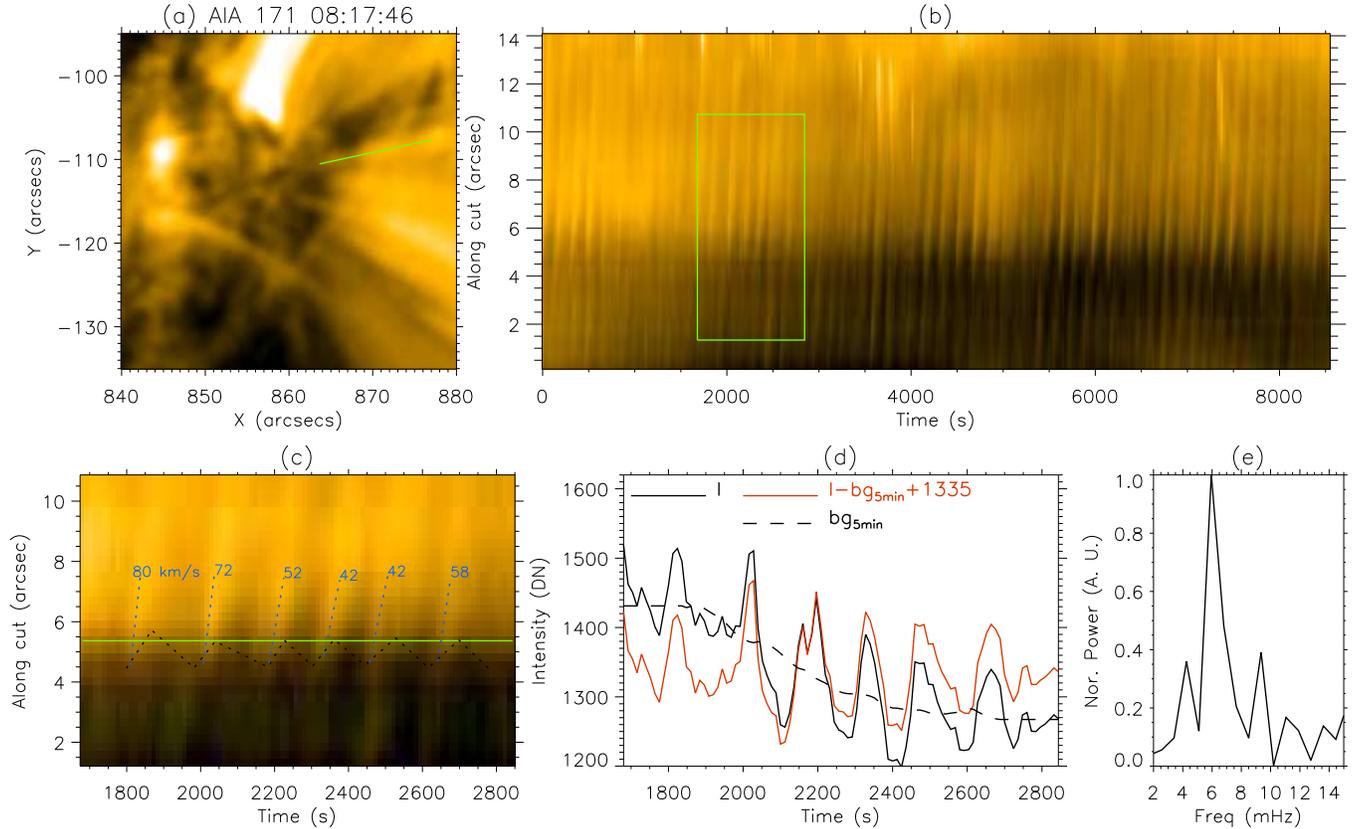}
\caption{
As Fig.\,\ref{fig:os1:171}, but for the case in \dataii.
}
\label{fig:os2:171}
\end{figure*}

In Fig.\,\ref{fig:os2:1400}, we show the periodicity analysis of the oscillating features seen in IRIS SJ 1400\,\AA\ observations.
The structure seems to be persistently oscillating as a whole in the direction perpendicular to its lengthp.
Fig.\,\ref{fig:os2:1400}b displays the time-distance image was obtained along the cut that was taken along the coronal loops (see Figs.\,\ref{f1} and \ref{fig:os2:171}) and marked in Fig.\,\ref{fig:os2:1400}a.
In the time-distance image, we can see clear sawtooth patterns, which are the manifestations of the oscillating feature.
The spatial amplitude of the oscillations is about 1\arcsec.
The apparent velocities are in the range of 11$\sim$14\,\kms\
while the oscillating feature is moving forward to the west,
and in the range of $4\sim8$\,\kms\ while the oscillating feature is moving backward to the east.
Considering the extending direction of the coronal loops,
the motion toward the west (east) can be considered to be upward (downward) propagation.
The speeds of the upward motion are comparable to the sound speed in the average chromosphere, and the downward motions are slower.
This indicates extra energy input in the upward motions.
In this set of data, the time-distance plot also presents a possible upward propagating component (see Fig.\,\ref{fig:os2:1400}c), although it is not as clear as that in \datai.

\par
The light curve taken from a given position along the cut (marked by the green line in Fig.\,\ref{fig:os2:1400}c)  is shown in Fig.\,\ref{fig:os2:1400}d.
The amplitudes of the intensity fluctuation are ranged from 45\% to 53\% of the background emissions.
Also, there is not any clear damping in the oscillation.
These large intensity amplitudes could be a hint of nonlinearity in the oscillations\,\citep{2014masu.book.....P,2014ApJ...786..137T}.

\par
The FFT power spectrum of the light curve is shown in Fig.\,\ref{fig:os2:1400}e.
We can see the main peak of the power spectrum at the frequency of 6.8\,mHz (i.e. 2.5 minutes in period).
This value suggests a connection between the oscillating feature and the prominent sunspot 3 minute oscillations.

\subsection{Upward propagating disturbances in SDO/AIA 171\,\AA\ observations}
\label{sect_coosc2}
The analysis results of the upward propagating disturbances along coronal loops rooted in the region are displayed in Fig.\,\ref{fig:os2:171}.
With the cut identical to that used in the analysis of the IRIS SJ images,
we produced the time-distance image of AIA 171\,\AA\ time series and shown in Fig.\,\ref{fig:os2:171}b.
The stripe-like patterns are the most clear features in the image.
These patterns are the response of the periodicity of the upward propagating disturbances along the cut (i.e. along the coronal loops).
Using the slopes of these stripe-like patterns,
the apparent velocities of the upward propagating disturbances are ranged from 42\,\kms\ to 80\,\kms\ (see Fig.\,\ref{fig:os2:171}c).
These values are much larger than that of the upward motions of the oscillating features seen in IRIS SJ (see the comparison of the tracks in Fig.\,\ref{fig:os2:171}c).
However, these values are significantly smaller than the sound speed in the corona.

\par
An example of light curves at a given point along the cut is shown in Fig.\,\ref{fig:os2:171}d.
The amplitudes of the intensity fluctuation are about 10\% of the background.
The FFT power spectrum of the light curve mainly peaks at the frequency of 6.8\,mHz,
as same as that found in the oscillating features observed by IRIS SJ.
This indicates that the oscillating features in the chromosphere/transition region are somehow connected to the upward propagating disturbances in the corona.

\begin{figure*}[!ht]
\centering
\includegraphics[trim=1cm -0.5cm 1cm 0.cm,width=0.9\textwidth]{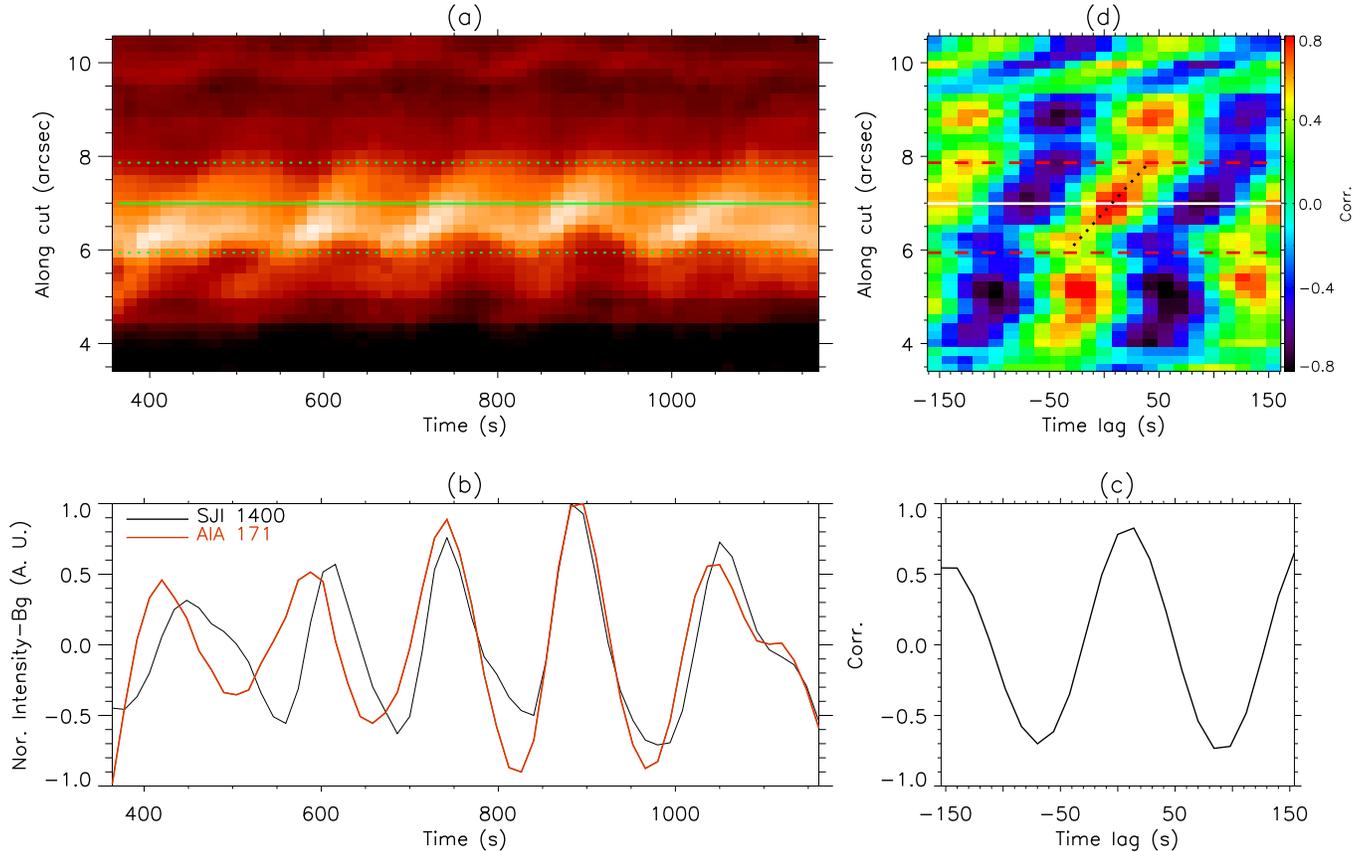}
\caption{
Analysis of time lags between the propagations of the upward motions of the oscillations and the upward propagating coronal disturbances in \datai\ along the cut shown in Figs.\,\ref{fig:os1:1400}\&\ref{fig:os1:171}.
In anticlockwise,
(a): The same time-distance image of the oscillating feature as shown in Fig.\,\ref{fig:os1:1400}c.
(b): The SJI 1400\,\AA\ and AIA 171\,\AA\ light curves taken at the locations of the cut denoted by the green solid line in panel (a) and white solid line in panel (d).
(c): Correlation coefficients between the SJI 1400\,\AA\ light curve and the AIA 171\,\AA\ light curve with various time lags.
(d): The map of correlation coefficients constructed along the cut with various time lags.
The region, where the oscillations of the features can be clearly identified, is marked by green dotted lines in panel (a) and red dashed lines in panel (d).
The black dotted line in panel (d) denotes the linear fit of the most probable time lags at different locations along the cut.
}
\label{fig:os1:lag}
\end{figure*}

\begin{figure*}[!ht]
\centering
\includegraphics[trim=1cm 0cm 1cm 0.cm,width=0.9\textwidth]{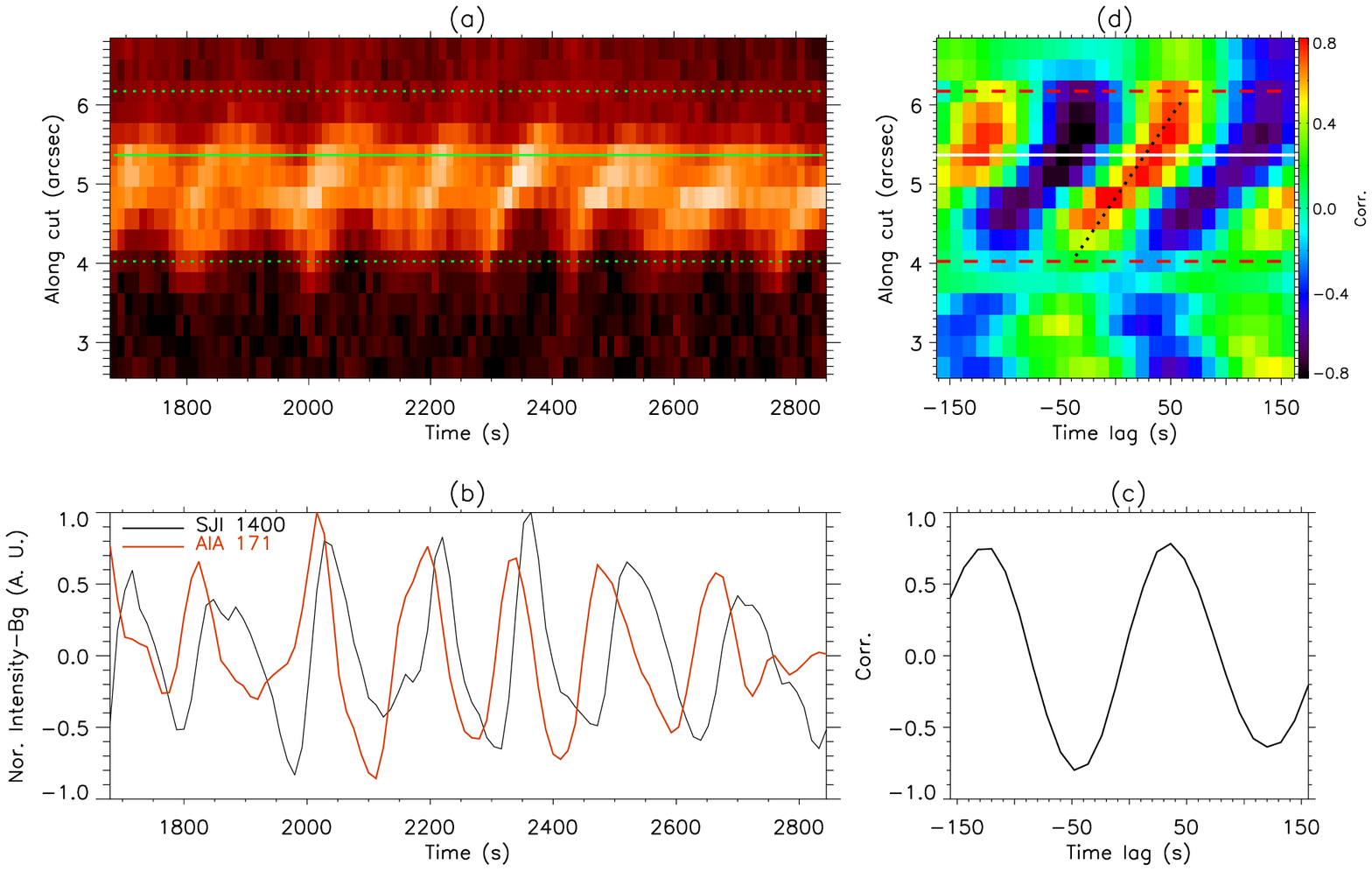}
\caption{
As Fig.\,\ref{fig:os1:lag}, but for the case observed in \dataii.
}
\label{fig:os2:lag}
\end{figure*}

\section{Time lags between the transition region oscillating features and the coronal disturbances}
\label{sect_link}
In the two sections above, we have presented observations of oscillations of the bright features in the upper-chromosphere/transition-region and upward propagating periodic disturbances along coronal loops rooted in the same region.
In this section, we will further investigate the time lags between propagations of the two periodic phenomena.

\par
By comparing the tracks of the two phenomena on the time-distance images (Figs.\,\ref{fig:os1:171}c and \ref{fig:os2:171}c),
we can see that the upward motions of the oscillating features appear earlier than the coronal disturbances, but moving slower.
We further determined the time lag between the two phenomena while propagating along the cuts (see Figs.\,\ref{fig:os1:lag} and \ref{fig:os2:lag}).
At any given position along the cut, the IRIS SJ 1400\,\AA\ and AIA 171\,\AA\ light curves (see e.g. Figs.\,\ref{fig:os1:lag}b and \ref{fig:os2:lag}b) are obtained,
and then we calculated the correlation coefficients between the two light curves by applying a variety of time lags to the AIA 171\,\AA\ light curve (see e.g. Figs.\,\ref{fig:os1:lag}c and \ref{fig:os2:lag}c).
A positive (negative) time lag is given while assuming that the AIA 171\,\AA\ disturbances are leading (lagging behind).
By applying the same procedures to all points along the cut, we can then construct a 2D map of the correlation coefficients (see Figs.\,\ref{fig:os1:lag}d and \ref{fig:os2:lag}d).

\par
The time lags at different locations of the cut are determined while maximum positive coefficients are obtained (see black dotted lines given in Figs.\,\ref{fig:os1:lag}d and \ref{fig:os2:lag}d).
The time lags are found to be negative at the bottom part of the cuts, but positive at the upper.
This indicates that the coronal disturbances lag the upward motion of the transition region features at the bottom of the loops (i.e. the cuts), but become leading after propagating awhile.
For the phenomena in \datai, the time lags change almost linearly from about $-30$\,s at the bottom to about $40$\,s after $\sim2$\arcsec\ propagation.
For that in \dataii, they vary almost linearly from about $-40$\,s at the bottom to about $60$\,s after $\sim2$\arcsec\ propagation.
These results are consistent with the velocity measurements from the time-distance images,
indicating that the propagations of the coronal disturbances and the oscillation of the transition region bright features are independent, i.e. without any interference.
The observed propagating coronal disturbances can be in the foreground (or background) of the oscillating features.

\begin{figure*}[!ht]
\centering
\includegraphics[trim=2cm 4cm 0.2cm 1cm,width=0.7\textwidth]{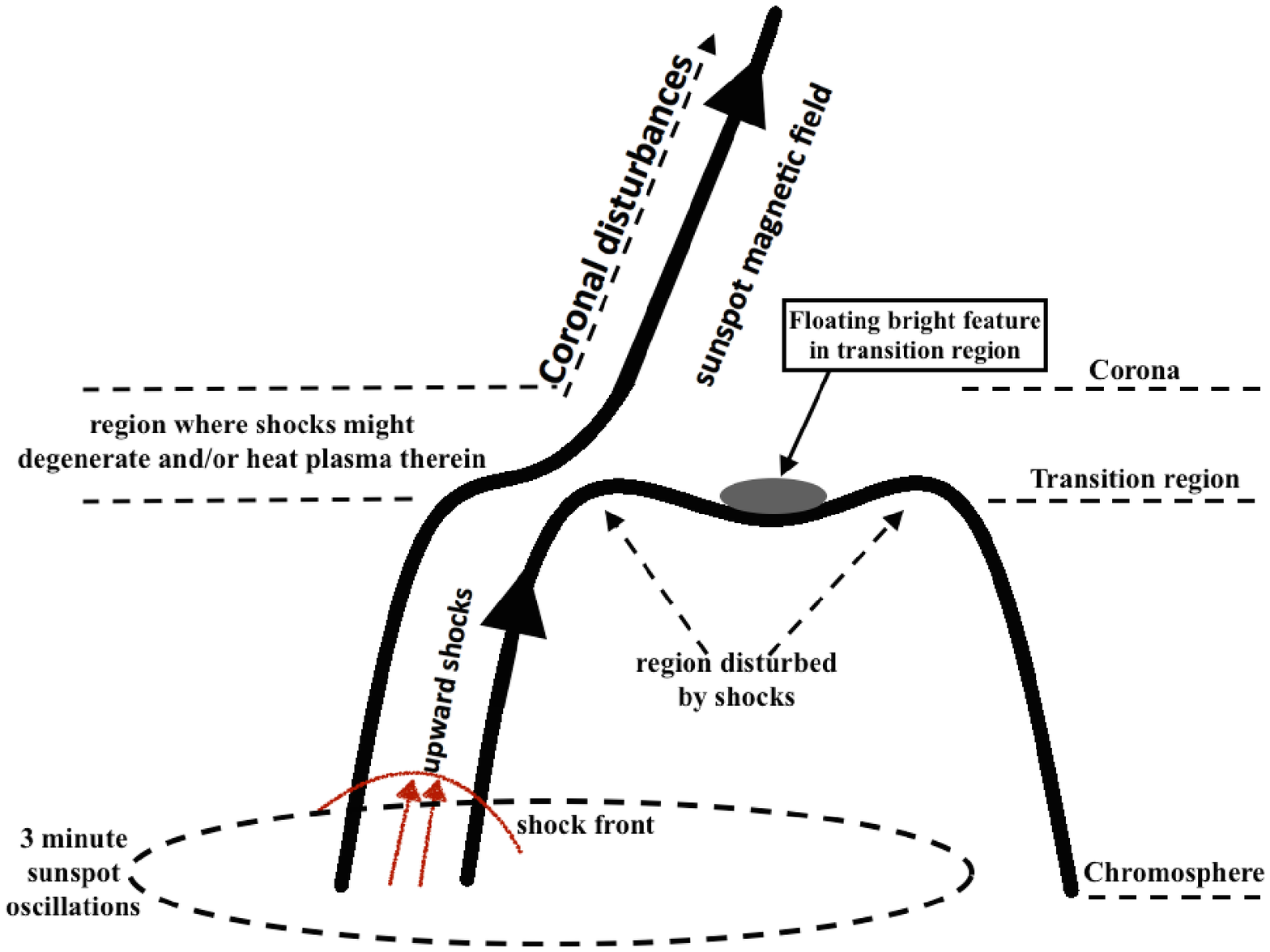}
\caption{
A cartoon scenario for the transition region oscillations and the upward propagating quasi-periodic coronal disturbances.
}
\label{fig:magmod}
\end{figure*}

\section{Discussions}
\label{sect_disc}
The observations of the upper-chromosphere/transition region oscillating features
   and the upward propagating coronal disturbances have been presented in the previous sections.
Base on the observations, in this section
   we will present our understanding as to the nature of and the connections between
   the transition region oscillations and the upward propagating coronal disturbances.

\par
The oscillatory motions of the bright featuresp in the transition region appear to be persistent,
which requires continuous energy input.
Based on the facts that these oscillations present some nonlinearity and
their periodicity is consistent with the omnipresent 3 minute sunspot oscillations,
a possible interpretation is that these oscillations are driven by the omnipresent 3 minute sunspot oscillations that might turn into shock in the upper solar atmosphere\,\citep{1984ApJ...277..874L,2014ApJ...786..137T,2015ApJ...804L..27Y,2017ApJ...838....2Z}.
Based on this scenario, a cartoon model is shown in Fig.\,\ref{fig:magmod}, in which the floating bright feature in the transition region is supported by dipped magnetic arches (represented by the field line on the right in Fig.\,\ref{fig:magmod}).
This is similar to that of typical prominences in the chromosphere\,\citep{2012LRSP....9....2A}.p
While shocks are propagating upward along the magnetic field, they can provide compressions and perturb the base of the feature and this further develops into persistent oscillations of the feature.
To some extend, these oscillations could be understand as forced harmonic oscillators that has been proposed to be a possible mechanism for oscillations in quiescent prominences\,\citep{1969SoPh....6...72K,2012LRSP....9....2A}.
Our interpretation is further discussed quantitively in the following.

\par
Let $P_{\rm nat}$ to denote the period of the oscillations that the structure undergoes upon an impulsive impact, and $\tau_{\rm nat}$ denotes the damping time.
Both $P_{\rm nat}$ and $\tau_{\rm nat}$ are solely determined by the physical parameters of the structure and its ambient.
Let $f(t)$, which is a function of time ($t$), to denote a cyclic force with a period of $P_{\rm drv}$.
With the idea of forced harmonic oscillators,
the transverse displacement $Y(t)$ of the structure as a whole can then be formally described as
\begin{eqnarray}
  \displaystyle
  \frac{\mathd^2 Y}{\mathd t^2} + 2\beta \frac{\mathd Y}{\mathd t} +
      \Omega^2 Y = f(t) ,
\label{eq_transdisp}
\end{eqnarray}
   where $\Omega = 2\pi/P_{\rm nat}$ and $\beta = 1/\tau_{\rm nat}$.
The periodicities of the oscillatory motions observed in both \datai\ and \dataii\ are close to 3 minutes despite that the structures in the two data sets show some evident difference.
A most likely cause is that this 3 minute periodicity is associated with the force (i.e. shocks in our interpretation) rather than the natural frequencies of the structures.
This is understandable given that the long-term behavior of the solution to Eq.~(\ref{eq_transdisp}) is solely determined by the period of the forcing unless $\beta = 0$.

\par
Another question is how to understand the monochromatic property of the oscillations.
To understand this issue, we assume that the forcing takes the form of
\begin{eqnarray}
   f(t) = \cos\left(\frac{\omega_{\rm drv} t}{2}\right),
   \label{eq_forcing}
\end{eqnarray}
where $\omega_{\rm drv} = 2\pi/P_{\rm drv}$, and $0\leq t\leq P_{\rm drv}$.
Now the solution to Eq.~(\ref{eq_transdisp}) for sufficently large $t$ is given by
\begin{eqnarray}
\displaystyle
   Y(t) = \sum_{n=1}^{\infty} C_n \cos\left(n \omega_{\rm drv}t - \phi_n\right) ,
\label{eq_disp_large_t}
\end{eqnarray}
    where
\begin{eqnarray}
\displaystyle
   C_n = -\frac{8}{\pi}\frac{n}{\left(4 n^2-1\right)}\frac{1}{\sqrt{\left(n^2 \omega_{\rm drv}^2 - \Omega^2\right)^2 + 4 n^2 \beta^2 \omega_{\rm drv}^2}}~.
\end{eqnarray}
Another set of constants $\phi_n$ is not relevant here.
The periodicity is only determined by the force frequency ($\omega_{\rm drv}$), while
the intrinsic frequency $\Omega$
only plays a role in determining the coefficients $C_n$.
Given the absence of clear signatures of harmonics with $n\ge 2$, some simple numerical evaluation yields that
   $|C_1|$ should be $\gtrsim 4|C_2|$.
Under the reasonable assumption that $\tau_{\rm nat} \gtrsim 2 P_{\rm nat}$,
   this requirement translates into $\Omega^2 \lesssim 2.2 \omega_{\rm drv}^2$, i.e.,
   $P_{\rm nat} \gtrsim P_{\rm drv}/1.5$.
With $P_{\rm drv} = 3~{\rm minutes}$, this means that the intrinsic oscillation period $P_{\rm nat}$
   needs to be longer than $\sim2$~minutes.
Could this condition be satisfied in reality?
Assuming that the restoring force of the oscillation is magnetic in nature, \citep{1969SoPh....6...72K} demonstrated that
   $P_{\rm nat} = L/v_{\rm A}$, where $L$ is the structure length and $v_{\rm A}$ is the Alfv\'en speed
   in the structure.
For the structure in DATA1, we find that the electron number density $n \approx 1.5\times 10^{10}$~cm$^{-3}$\,\citep{2016ApJ...829L..30H},
and the length of this structure is found to be roughly $1.5\times10^4$\,km.
It then follows that the magnetic field strength in this structure should be $\lesssim 12$~Gs for $P_{\rm nat}$ to be longer than $\sim 2$~mins.
This value is well consistent with the measured magnetic fields in quiescent prominences\,\citep{2010SSRv..151..333M}.
In fact, if gravity or gas pressure is further incorporated into the restoring force, $P_{\rm nat}$ can be much longer\,\citep[e.g.][]
{1974A&A....31..189K, 2009A&A...497..521A, 2016A&A...590A.120K}.

\par
The upward quasi-monochromatic disturbances seen in AIA 171\,\AA\
   can also be interpreted as the coronal response to the recurrent shocks,
   provided that the shocks remain coherent over a distance comparable to the size
   of the umbra (see Fig.\,\ref{fig:magmod}).
Upward propagating quasi-periodic coronal disturbances are abundant in coronal holes and
   fan loops with periods from about one minute to more than 20 minutes\,\citep[see e.g.][etc.]{1998ApJ...501L.217D,2003A&A...404L..37M,2005LRSP....2....3N,2007Sci...318.1574D,2009A&A...503L..25W,2010ApJ...722.1013D,2011ApJ...736..130T,2012ApJ...759..144T}.
It has been suggested that these coronal disturbances
   are slow magnetic-acoustic waves\,\citep[e.g.][]{2009A&A...503L..25W} or recurring upflows\,\citep[e.g.][]{2011ApJ...738...18T,2011ApJ...736..130T,2012ApJ...759..144T}.
In the present observations, the periodicity of the upward propagating coronal disturbances
   strongly suggests a connection to the 3 minute sunspot oscillations.
Based on the shock scenario discussed above,
the upward propagating coronal disturbances could be recurrent plasma flows that were heated and driven by shocks, or
alternatively they could be responses of degenerated shocks that become slow magnetic-acoustic waves after heating the plasma in the coronal loops at their transition-region bases (represented by the field line on the left in Fig.\,\ref{fig:magmod}).
Since shocks could be one of the most plausible sources of spicules\,\citep[see][and references therein]{2000SoPh..196...79S},
this discussion is consistent with the findings of their strong connections to spicules\,\citep[e.g.][]{2011Sci...331...55D,2015ApJ...809L..17J,2016RAA....16...93J}.

\par
In this section, we have present our scenario to understand the oscillating features in the transition region and the coronal disturbances.
In this scenario, 3 minute oscillations are coherent over a large spatial scale from chromosphere to transition region.
The upward propagating waves associated with these 3 minute oscillations present as shocks below transition region, and appear to be slow magneto-acoustic waves or plasma flows in the corona.
The shocks have been heating plasmas and degenerated in their upward propagating trajectory.
This provides an approach for energy transporting from the  the chromosphere to the corona.
Although our understandings to the phenomena observed here appear to be the most plausible, some other mechanisms cannot be excluded.


\section{Conclusions}
\label{sect_concl}
With high-resolution IRIS SJ 1400\,\AA\ data, we observed oscillations of the transition region bright features floating above the sunspot umbrae.
The oscillations are persistent motionsp with period of about 3 minutes.
This suggests their strong connections to the omnipresent 3 minute oscillations in sunspots.
The spatial displacement (i.e. amplitude) of the oscillation of the feature is $\lesssim$1\arcsec.
The apparent velocities of the oscillating motions are about 10\,\kms, and the velocities of the downward motions are slightly smaller than that of the upward.
The intensity fluctuations of the oscillations show amplitudes of more than 24\% of the background emissions,
indicating of some nonlinear effects.
Furthermore, the oscillations did not show any obvious damping,
and this requires continuous energy inputs during our observations.

\par
In the coronal loops rooted in the same regions as the oscillating features,
we observed upward propagating quasi-periodic disturbances with AIA 171\,\AA\ data.
The coronal disturbances have exactly the same periodicity ($\sim$3 minutes) as that of the oscillations of the bright features in the transition region.
This suggest that both are associated with the sunspot 3 minute oscillations.
The peaks of the intensity fluctuations of the coronal disturbances take only 10$\sim$15\% of the background.
The apparent upward propagating velocities measured in AIA\,171\,\AA\ images are about 30\,\kms\ in the cases observed in \datai\ and in the range of 40$\sim$80\,\kms\ in the cases seen in \dataii.
These values are much smaller than the sound speed of average corona.

\par
Based on our observations,
we suggest that the persistent oscillations of the bright features in the transition region are possibly powered by upward propagating shocks.
The persistent propagations of shocks from lower solar atmosphere can provide continuous energy inputs that propel the oscillations.
The upward propagating quasi-periodic coronal disturbances could be recurrent plasma flows driven by the shocks in the lower solar atmosphere.
Alternatively, the upward propagating quasi-periodic coronal disturbances might be responses of degenerated shocks that become slow magnetic-acoustic waves after heating the plasma in the coronal loops at their transition-region bases.

\acknowledgments
{\it Acknowledgments:}
We greatly acknowledge the anonymous referee for the comments.
We thank Dr. Shuhong Yang for many discussions, suggestions and comments, and Dr. Xiaoshuai Zhu for helping to handle the HMI  full-disk vector magnetic field.
This research is supported by
National Natural Science Foundation of China (41474150, 41627806, 41474149, 41404135 and 41274178).
ZH thanks the China Postdoctoral Science Foundation and the Young Scholar Program of Shandong University, Weihai (2017WHWLJH07).
IRIS is a NASA small explorer mission developed and operated by LMSAL with mission operations
executed at NASA Ames Research center and major contributions to downlink communications
funded by the Norwegian Space Center through an ESA PRODEX contract.

\bibliographystyle{aasjournal}
\bibliography{bibliography2}

\end{document}